\documentclass[aps,prc,nofootinbib,superscriptaddress,onecolumn,11pt,floatfix]{revtex4-2}
\usepackage{amsmath}
\usepackage{amssymb}

\usepackage{hyperref}

\usepackage{liealg}
\usepackage{writelabel}

\usepackage{xcolor}

\begin{document}

\title{One-body and two-body density matrix elements in a symplectic many-body basis}

\author{Jakub Herko}
\affiliation{Department of Physics and Astronomy, University of Notre Dame, Notre Dame, Indiana 46556-5670, USA}
\affiliation{TRIUMF, Vancouver, British Columbia V6T 2A3, Canada}

\author{Mark A. Caprio}
\affiliation{Department of Physics and Astronomy, University of Notre Dame, Notre Dame, Indiana 46556-5670, USA}

\date{\today}

\begin{abstract}
The symplectic no-core configuration interaction (SpNCCI) framework is an \textit{ab initio} many-body method for nuclear structure which makes use of the approximate symplectic symmetry of nuclei by appropriate choice of many-body basis states. In this paper we derive recurrence relations allowing for calculation of one-body and two-body density matrix elements between the SpNCCI basis states. Availability of these matrix elements allows for integration of the SpNCCI framework with other modern many-body methods and for calculation of matrix elements of any one-body or two-body operator.
\end{abstract}
\maketitle

\writelabel{part1:sec:u3}{II}
\writelabel{part1:section:u3-many-body}{III}
\writelabel{part1:subsection:su3-many-body}{III~A}
\writelabel{part1:subsection:su3basis-me}{III~B}
\writelabel{part1:section:sp3r}{IV}
\writelabel{part1:subsection:sp3r_alg}{IV~A}
\writelabel{part1:subsection:sp3r_irrep}{IV~B}
\writelabel{part1:subsection:sp3r-generator-rme}{IV~C}
\writelabel{part1:subsection:sp3r-many-body}{V}
\writelabel{part1:section:recurrence}{VI}
\writelabel{part1:subsection:peelA}{VI~A}
\writelabel{part1:subsection:commuting}{VI~B}
\writelabel{part1:subsection:evaluate:AUcom}{VI~C}
\writelabel{part1:sec:app-commutator}{B}
\writelabel{part1:sec:app-coupled-commutator-unit-tensors}{C}

\writelabel{part1:eqn:eq:Q_mass}{1}
\writelabel{part1:eq:boson-tensors}{17}
\writelabel{part1:sp_generators_canonical}{44}
\writelabel{part1:sp_generators_boson}{45}
\writelabel{part1:tensor_generators}{46}
\writelabel{part1:raising_polynomial_recurrence}{48}
\writelabel{part1:n_boson}{49}
\writelabel{part1:boson_intrinsic}{60}
\writelabel{part1:polynomial_expansion_manybody_basis}{62}
\writelabel{part1:intrinsic-sp3r-gen}{61}
\writelabel{part1:cfp1}{71}
\writelabel{part1:eq:relative_unit_tensor_u3s_rme_overlap}{73}
\writelabel{part1:eq:unit_tensor_mb_rme1_generic}{77}
\writelabel{part1:eq:recurrence_final}{84}
\writelabel{part1:eqn:coupled-commutator-su3-product}{B9}
\writelabel{part1:eqn:coupled-commutator-su3-su2-difference}{B11}

\writelabel{part1:fig:sp_laddering}{1}

\section{Introduction}

Recent progress in \textit{ab initio} nuclear many-body theory has extended the reach of \textit{ab initio} methods to medium-mass~\cite{morris2018,gysbers2019,yao2020,stroberg2021,hergert2020,belley2021,hu2024} and even heavy~\cite{hu2022} nuclei. However, description of nuclear collective motion within the \textit{ab initio} methods still remains a challenge, because it requires model spaces of sizes which often exceed current computational limits. This challenge can be addressed by using an approximate symmetry of nuclei. By approximate symmetry of nuclei we mean the fact that nuclear wave functions contain dominant contributions of a few irreducible representations (irreps) of a symmetry group. In symmetry-based approaches we reorganize the model space into irreps of the symmetry group and try to truncate the model space by keeping only irreps that are important for description of nuclear states, thereby reducing the size of the model space.

It has been shown that the non-compact symplectic group $\grpsptr$\footnote{The group is sometimes denoted by $\grpspr{6}$ in the literature.}~\cite{rosensteel1977i,rowe1985jpa} is a group describing an approximate symmetry of nuclei~\cite{dytrych2007,dytrych2007prc,dytrych2020,launey2020}. This approximate symmetry is used in the \textit{ab initio} symplectic no-core configuration interaction (SpNCCI) framework~\cite{mccoy2018dis,mccoy2018,part1}, in which we solve the Schr\"odinger equation as an eigenvalue problem for the Hamiltonian matrix constructed in a symplectic many-body basis.

To construct the Hamiltonian matrix in the symplectic many-body basis, we need to calculate matrix elements of the kinetic energy and of the internucleon interaction. The kinetic energy is a generator of $\grpsptr$, and mathematical tools for calculation of matrix elements of the generators of $\grpsptr$ in a symplectic basis have long been available~\cite{rowe1984,rowe19842,rowe2016,part1}. Also, for the calculation of the matrix elements of a two-body interaction, a method based on a recurrence relation has been developed~\cite{reske1984,suzuki1986,mccoy2018dis,part1}. This method more generally can be used to calculate matrix elements of any intrinsic two-body operator.\footnote{By an intrinsic operator, we mean a Galilean invariant many-body operator.  Such an operator can depend upon the coordinates and momenta of the nucleons relative to the center or mass, that is, in the translational intrinsic frame, and upon the spin degrees of freedom, but not on the coordinate or momentum of the center of mass itself.}  Such operators include not only two-body interactions, but also, \textit{e.g.}, the magnetic dipole and electric quadrupole operators in the impulse approximation (when properly defined relative to the center of mass of the nuclear system), thereby providing the corresponding electromagnetic moments and transition probabilities, or, more generally, any operator which is bilinear in the coordinates and momenta of the nucleons~\cite{caprio2020}.

However, in some many-body methods, we furthermore need one-body or two-body density matrix elements. Examples are the normal ordered two-body approximation of a three-body interaction~\cite{roth2012,gebrerufael2016,hebeler2023} and the in-medium similarity renormalization group approach~\cite{tsukiyama2011,bogner2014,stroberg2016,hergert2016}, which use Wick’s theorem, yielding contractions of creation and annihilation operators, which are given by one-body density matrix elements. Another example is the resonating-group method~\cite{tang1978,quaglioni2008,quaglioni2009,navratil2016}, which organizes particles into clusters and involves integral kernels calculated from one-body and two-body density matrix elements. To integrate the SpNCCI framework with these methods, we need one-body and two-body density matrix elements between symplectic many-body basis states.

These matrix elements can also be used to calculate matrix elements of any one-body or two-body operator, provided the one-body or two-body matrix elements of the operator are available. For example, matrix elements of one-body operators are needed to calculate electron scattering form factors, which carry detailed information about nuclear structure~\cite{escher1997}.

In this paper we derive recurrence relations allowing for calculation of one-body and two-body density matrix elements between symplectic many-body basis states. Note that, to properly address the center-of-mass degree of freedom, we consider the intrinsic realization of the generators of $\grpsptr$, which may be defined either in terms of the Jacobi coordinates and momenta of the particles or, equivalently, in terms of coordinates and momenta with respect to the center-of-mass frame, not simply the generators defined in terms of coordinates and momenta in the laboratory frame. This yields symplectic many-body basis states which are free of spurious center-of-mass excitations. Although a recurrence relation for the one-body density matrix elements was previously reported in Refs.~\cite{escher1997,escher1998}, this derivation was based on an invalid assumption that certain center-of-mass correction terms should vanish.

We first review relevant aspects of the SpNCCI framework, in Sec.~\ref{sec:background}, before proceeding with the derivations of the recurrence relations themselves, in Sec.~\ref{4.recu}. These recurrence relations were initially presented in Ref.~\cite{herko2024}.

\section{Background}
\label{sec:background}

In this section, we review aspects of the SpNCCI framework relevant to the derivation of recurrence relations for the density matrix elements.  First, we review the symplectic raising operator, which is used to construct the basis of the SpNCCI model space, and express it in terms of fermion creation and annihilation operators in a form needed for the derivation of the recurrence relations for the density matrix elements (Sec.~\ref{sec:background:raising}). Then we review the construction and labeling scheme for the SpNCCI basis states (Sec.~\ref{sec:background:basis}). Finally, we review those aspects of the recurrence for matrix elements of a many-body operator in the SpNCCI basis which are common for intrinsic operators and operators dependent on laboratory-frame single-particle coordinates (Sec.~\ref{sec:background:recursive}).  For the reader's convenience, we provide cross-references to relevant sections and equations of Ref.~\cite{part1}, where much of this background is described in greater detail.

\subsection{Symplectic raising operator}
\label{sec:background:raising}

The significance of the symplectic group $\grpsptr$ for description of an $A$-particle system stems from the physical significance of its generators, reviewed in Sec.~\ref{part1:subsection:sp3r_alg} of Ref.~\cite{part1}. The 21 generators of $\grpsptr$ can be represented by bilinear products of the spatial coordinates and momenta of the particles~\cite{rowe1985,part1}, as given by~(\ref{part1:sp_generators_canonical}) of Ref.~\cite{part1}. These include physically relevant operators such as the many-body kinetic energy and harmonic oscillator (HO) Hamiltonian as well as orbital angular momentum and quadrupole operators and the monopole (or squared radius) operator. Whereas the realization of the generators of $\grpsptr$ in terms of the bilinear products of the canonical coordinates provides the most direct connection to observables, for calculations in the SpNCCI framework, with a Hilbert space consisting of linear combinations of HO configurations, it is more convenient to realize the generators of $\grpsptr$ in terms of the HO raising and lowering operators~\cite{quesne1971,part1}, as given by~(\ref{part1:sp_generators_boson}) of Ref.~\cite{part1}. The generators of $\grpsptr$ include generators of Elliott's $\grpu{3}$ and $\grpsu{3}$~\cite{elliott1958}, which thus are subgroups of $\grpsptr$.

For calculations in a symmetry-adapted basis it is useful to express the symplectic generators as $\grpsu{3}$ irreducible tensor operators~\cite{rosensteel1992:sp3r-tensors-gtssnp91,escher1997,escher1998,prc-65-2002-054309-Escher}, so that group-theoretical tools such as the $\grpsu{3}$ Wigner-Eckart theorem can be used to simplify the calculations. The HO raising and lowering operators can be recognized as $\grpsu{3}$ irreducible tensor operators $c^{\dagger(1,0)}$ and $c^{(0,1)}$, transforming according to the $(1,0)$ and $(0,1)$ irreps of $\grpsu{3}$, as given by~(\ref{part1:eq:boson-tensors}) of Ref.~\cite{part1}. Then the HO raising and lowering operators, thus expressed as $\grpsu{3}$ irreducible tensor operators, can be coupled to obtain the symplectic generators in the form of $\grpsu{3}$ irreducible tensor operators, as given by~(\ref{part1:tensor_generators}) of Ref.~\cite{part1}. In particular, for the symplectic raising operator we have
\begin{equation}\label{noninva}
A^{(2,0)}=\frac{1}{\sqrt{2}}\sum_{s=1}^{A}\left[c^{\dagger(1,0)}_s\times c^{\dagger(1,0)}_s\right]^{(2,0)}.
\end{equation}

The symplectic raising operator~(\ref{noninva}) thus defined, in terms of positions and momenta of the particles in the laboratory frame, is not Galilean invariant, and therefore gives rise to spurious center-of-mass excitations. This problem can be solved by instead defining the symplectic raising operator in terms of the $A-1$ Jacobi coordinates and momenta of the $A$ particles~\cite{escher1997,escher1998}, or in terms of positions and momenta defined with respect to the center-of-mass frame~\cite{rosensteel1980npa}.

The intrinsic symplectic raising operator can be expressed in terms of the laboratory-frame symplectic raising operator $A^{(2,0)}$ of~(\ref{noninva}), from which we subtract a center-of-mass correction. This yields the intrinsic symplectic raising operator $A'^{(2,0)}$ as given in~(\ref{part1:intrinsic-sp3r-gen}) of Ref.~\cite{part1}:
\begin{equation}\label{syraop}
A'^{(2,0)}=A^{(2,0)}-A^{(2,0)}_{\rm cm},
\end{equation}
where
\begin{equation}\label{4.8}
A^{(2,0)}_{\rm cm}=\frac{1}{\sqrt{2}}\left[c^{\dagger(1,0)}_{\rm cm}\times c^{\dagger(1,0)}_{\rm cm}\right]^{(2,0)}
\end{equation}
is the center-of-mass realization of the symplectic raising operator, defined in terms of the center-of-mass HO raising operator
\begin{equation}\label{kra}
c^{\dagger(1,0)}_{\rm cm}=\frac{1}{\sqrt{A}}\sum_{t=1}^{A}c^{\dagger(1,0)}_{t}
\end{equation}
given in~(\ref{part1:boson_intrinsic}) of Ref.~\cite{part1}.

The intrinsic symplectic raising operator $A'^{(2,0)}$ is used to construct the SpNCCI many-body basis states, as described in Sec.~\ref{part1:subsection:sp3r-many-body} of Ref.~\cite{part1}. Whereas the laboratory-frame symplectic raising operator $A^{(2,0)}$ is a one-body operator, the intrinsic operator contains a two-body part originating from the center-of-mass correction. This has important consequences for calculation of density matrix elements between symplectic many-body basis states.

To calculate the density matrix elements between the symplectic many-body basis states, we must take the symplectic raising operator, as given in terms of the HO raising operator above, and reexpress it in second-quantized form, namely, in terms of the fermion creation and annihilation operators for fermions in HO single-particle states.  The creation operators $a_{Nlm_lm_s}^{\dagger}$ of fermions with spin $\frac{1}{2}$ in the spherical HO basis ($N$ is the HO shell index, while $m_l$ and $m_s$ are the projection quantum numbers of the orbital angular momentum $l$ and spin, respectively) are irreducible tensor operators $a_{lm_lm_s}^{\dagger(N,0)\frac{1}{2}}$ with respect to the spatial $\grpsu{3}$ and the intrinsic spin $\grpsu[S]{2}$ symmetries with tensor characters ($N,0$) and $\frac{1}{2}$, respectively (see Sec.~\ref{part1:subsection:su3basis-me} of Ref.~\cite{part1})~\cite{draayer1973}.  We label these operators by their combined $\grpsu{3}\times\grpsu[S]{2}$ tensor character $(\lambda,\mu)S=(N,0)\tfrac12$.  The corresponding annihilation operators $a_{Nlm_lm_s}$, as obtained by simple Hermitian conjugation of the creation operators, do not form a standard basis for an $\grpsu{3}\times\grpsu[S]{2}$ irreducible tensor.  Therefore, one needs to introduce the related operators
\begin{equation}\label{4.til}
\tilde{a}_{lm_lm_s}^{(0,N)\frac{1}{2}}=(-1)^{N+l-m_l+\frac{1}{2}-m_s}a_{Nl-m_l-m_s},
\end{equation}
which do transform as a standard basis of tensor character $(0,N)\frac{1}{2}$.

The fermionic realization of the intrinsic symplectic raising operator is given in Refs.~\cite{escher1997,escher1998}, where it is expressed as a sum of a one-body operator and a two-body operator.  However, for the calculation of the density matrix elements, where the center-of-mass HO raising operator will play an important role, we will find it convenient to instead express the intrinsic symplectic raising operator as in~(\ref{syraop}), where $A^{(2,0)}$ is a one-body operator and $A^{(2,0)}_{\rm cm}$, given by~(\ref{4.8}), is a coupled product of two one-body operators, which contains both one-body and two-body parts. The fermion realization of the laboratory-frame symplectic raising operator $A^{(2,0)}$ in the HO single-particle basis is provided in Refs.~\cite{escher1997,escher1998}, and that of the center-of-mass HO raising operator $c^{\dagger(1,0)}_{\rm cm}$ can be derived similarly (\textit{c.f.}\ Ref.~\cite{luo2013}). The results are
\begin{align}
A^{(2,0)0}&=\sum_{\nu}\sqrt{\frac{(\nu+1)(\nu+2)(\nu+3)(\nu+4)}{12}}\left[a^{\dagger}_{\nu+2}\times\tilde{a}_{\nu}\right]^{(2,0)0},\label{4.6}\\
c^{\dagger(1,0)0}_{\rm cm}&=\sum_{\nu}\sqrt{\frac{(\nu+1)(\nu+2)(\nu+3)}{3A}}\left[a^{\dagger}_{\nu+1}\times\tilde{a}_{\nu}\right]^{(1,0)0},\label{Bdag}
\end{align}
where we write $a^{\dagger}_{\nu}$ and $\tilde{a}_{\nu}$ instead of $a^{\dagger(\nu,0)\frac{1}{2}}$ and $\tilde{a}^{(0,\nu)\frac{1}{2}}$, for brevity of notation. Note that the operators $A^{(2,0)0}$ and $c^{\dagger(1,0)0}$, appearing in~(\ref{4.6}) and~(\ref{Bdag}), act only on the spatial degrees of freedom, and therefore must be scalar with respect to spin, as we now explicitly indicate in their $\grpsu{3}\times\grpsu[S]{2}$ tensor labels.

\subsection{Basis of the SpNCCI model space}
\label{sec:background:basis}

In the SpNCCI framework, the many-body model space is decomposed into $\grpsptr$ irreps, as described in Sec.~\ref{part1:subsection:sp3r-many-body} of Ref.~\cite{part1}. Here we briefly review the construction and labeling scheme, both of basis states for $\grpsptr$ irreps generally, and of the SpNCCI many-body basis states in particular. More details can be found in Secs.~\ref{part1:subsection:sp3r_irrep} and~\ref{part1:subsection:sp3r-many-body} of Ref.~\cite{part1}.

Since the group $\grpu{3}$ is a subgroup of $\grpsptr$, we further decompose each $\grpsptr$ irrep into $\grpu{3}$ irreps. The symplectic generators include the generators of $\grpu{3}$, which conserve the number of HO quanta and do not connect different $\grpu{3}$ irreps, and the symplectic raising and lowering operators, which change the number of HO quanta by two and connect different $\grpu{3}$ irreps. Many-body basis states of a symplectic irrep are constructed by starting from a $\grpu{3}$ irrep $\sigma$, defined by the property that its states are annihilated by the symplectic lowering operator, and then acting repeatedly with the symplectic raising operator, which increases the number of HO quanta in increments of two.  (The action of the symplectic raising operator can be repeated without limit, yielding an infinite number of states.\footnote{Non-compact groups such as $\grpsptr$ have infinite-dimensional representations.}) The states within the starting $\grpu{3}$ irrep have the lowest number of HO quanta within the symplectic irrep. The starting irrep is correspondingly referred to as the lowest grade irrep (LGI).\footnote{The grade here refers to the $\grpu{1}$ label representing the number of oscillator quanta.  The term symplectic bandhead is also used for the LGI in the literature.}  Its $\grpu{3}$ label $\sigma$ is also used as the $\grpsptr$ label, that is, to label the whole symplectic irrep.

In terms of the $\grpu{3}$ many-body basis~\cite{draayer2012,dytrych2013,dytrych2016,langr20192,part1}, described in Sec.~\ref{part1:section:u3-many-body} of Ref.~\cite{part1}, the LGIs can be obtained explicitly by solving for the null space of the symplectic lowering operator, represented as a matrix in this basis.\footnote{All $\grpu{3}$ irreps with fewer than $2$ HO excitation quanta, namely, those with either $0$ or $1$ excitation quanta, are annihilated by the symplectic lowering operator, and are thus automatically LGIs.} To ensure that the LGIs are free of spurious center-of-mass excitations~\cite{luo2013}, we solve for the simultaneous null space of the symplectic lowering operator and of the number operator for HO quanta in the center-of-mass motion. Details of this procedure can be found in Sec.~\ref{part1:subsection:sp3r-many-body} of Ref.~\cite{part1}.

The symplectic raising operator is then applied to the LGI specifically by acting on that $\grpu{3}$ irrep with a polynomial $P^n(A')$ in the symplectic raising operator, with the factors symmetrically coupled to yield the $\grpu{3}$ irrep $n$,\footnote{Since the components of the symplectic raising operator commute, only the symmetrically coupled polynomials are nonvanishing~\cite{rosensteel1980}.} and the LGI and raising polynomial are then coupled to yield $\grpu{3}$ irreps $\omega$ with increased number of oscillator quanta $N_{\omega}=N_{\sigma}+N_n$.  The resulting organization of an $\grpsptr$ irrep into $\grpu{3}$ irreps, differing by HO quanta in steps of two, is illustrated in Fig.~\ref{part1:fig:sp_laddering} of Ref.~\cite{part1}.  The many-body space can be organized into $\grpsptr$ irreps and then, within these, into $\grpu{3}$ irreps, which are linear combinations of the original $\grpu{3}$ many-body basis irreps.  For any given number of HO quanta, some of these $\grpu{3}$ irreps are members of an $\grpsptr$ irrep obtained by laddering from an LGI with a lower number of HO quanta, while the remaining $\grpu{3}$ irreps form the LGIs of their own $\grpsptr$ irreps.

The many-body basis states of a symplectic irrep are free of spurious center-of-mass excitations as long as they are constructed starting from an LGI which is free of center-of-mass excitations and then are obtained from this LGI by laddering using the intrinsic symplectic raising operator. Furthermore, the many-body basis states are fully antisymmetric with respect to exchange of identical fermions, because the symplectic raising operators preserve the antisymmetry of the LGIs.

Since the group SO(3) is a subgroup of $\grpu{3}$, we further decompose each $\grpu{3}$ irrep into SO(3) irreps labeled by orbital angular momentum $L$. Since the symplectic generators act only on the nucleon spatial coordinates, leaving the spin degree of freedom unaffected, each symplectic irrep carries a definite total intrinsic spin $S$, which is coupled with $L$ to give the total angular momentum $J$, with projection $M$. Thus, the many-body basis states are labeled by $|\gamma\sigma\upsilon\omega\kappa L\Sigma SJM\rangle$, where $\gamma$ and $\Sigma$ denote labels distinguishing different $\grpsptr$ irreps with the same LGI $\grpu{3}\times \grpsu[S]{2}$ labels $\sigma S$,\footnote{The specific definition of $\Sigma$ depends on whether we use the proton-neutron coupling scheme or the isospin coupling scheme to construct the LGIs. In the former scheme, $\Sigma$ refers to proton and neutron spin quantum numbers, whereas in the latter scheme it involves SU(4) quantum numbers (see Sec.~\ref{part1:subsection:su3-many-body} of Ref.~\cite{part1} for details).} $\upsilon$ is a branching multiplicity index from $\grpsptr$ to $\grpu{3}$, which distinguishes different $\grpu{3}$ irreps with the same labels $\omega$ within an $\grpsptr$ irrep, and $\kappa$ is an inner multiplicity index distinguishing SO(3) irreps with the same label $L$ within an $\grpu{3}$ irrep.

The raising polynomial $P^n(A')$ can be represented recursively~\cite{suzuki1986}, in terms of raising polynomials of lower order, using~(\ref{part1:raising_polynomial_recurrence}) of Ref.~\cite{part1}.  This allows for peeling off a symplectic raising operator from a symplectic many-body basis state using~(\ref{part1:cfp1}) of Ref.~\cite{part1}, which in turn allows for recursive construction of the many-body basis states of a symplectic irrep starting from the LGI.

However, the many-body states thus defined do not form an orthonormal basis.  Specifically, the $\grpu{3}$ irreps with the same labels $\omega$, but which are obtained with different raising polynomials (with different $n$), or which arise as different instances of the irrep $\omega$ occuring in the Kronecker product of $\sigma$ and $n$ (distinguished by the outer multiplicity index $\rho$), are not, in general, orthogonal. That is, the symplectic many-body basis states $|\gamma\sigma n\rho\omega\kappa L\Sigma SJM\rangle$ constructed from the LGIs by action of the symplectic raising operators in the aforementioned way are not orthogonal with respect to the labels $n\rho$.  In certain restricted cases, the set thus generated may also be overcomplete.  The orthonormal basis states $|\gamma\sigma\upsilon\omega\kappa L\Sigma SJM\rangle$ are obtained from the non-orthogonal (and possibly overcomplete) states via the $K$-matrix expansion given by~(\ref{part1:polynomial_expansion_manybody_basis}) of Ref.~\cite{part1}, which reads
\begin{equation}
|\gamma\sigma\upsilon\omega\kappa L\Sigma SJM\rangle=\sum_{n\rho}(K^{\omega}_{\sigma})^{-1}_{\upsilon[n\rho]}|\gamma\sigma n\rho\omega\kappa L\Sigma SJM\rangle,
\end{equation}
where $K^{\omega}_{\sigma}$ is the $K$ matrix~\cite{rowe1984,rowe1995}, reviewed in Sec.~\ref{part1:subsection:sp3r_irrep} of Ref.~\cite{part1}.  Vector coherent state (VCS) theory~\cite{rowe1984,rowe19842,hecht1987,rowe1995,rowe2016} provides a recurrent method for the calculation of the $K$ matrices~\cite{rowe1984,hecht1987,rowe1988,rowe1995,rowe2016}, as reviewed in Sec.~\ref{part1:subsection:sp3r_irrep} of Ref.~\cite{part1}.

\subsection{Recursive calculation of matrix elements in the symplectic many-body basis}
\label{sec:background:recursive}

In principle, one can calculate matrix elements, in the SpNCCI basis, of a many-body operator by expanding all the SpNCCI basis states in terms of the U(3) many-body basis.\footnote{After expanding the SpNCCI basis states in terms of the $\grpu{3}$ many-body basis, one can calculate the matrix elements using techniques developed in Ref.~\cite{bahri1994}.} This can be done by first expanding the LGIs and then using the recursive SpNCCI basis construction, as described in Sec.~\ref{sec:background:basis}. However, such calculation of matrix elements is computationally very demanding. A more feasible approach is to use the $\grpu{3}$ many-body basis to calculate only the matrix elements (of some set of operators of interest) between the LGIs and then calculate the remaining matrix elements recursively~\cite{reske1984,suzuki1986,part1}. Here we review those aspects of the recursive calculation which are applicable regardless of whether the operators involved are intrinsic or depend upon the single-particle coordinates (details can be found in Secs.~\ref{part1:subsection:peelA} and~\ref{part1:subsection:commuting} of Ref.~\cite{part1}).

The recurrence relation relates $\grpsu{3}\times\grpsu[S]{2}$ reduced matrix elements (RMEs) of a restricted family of $\grpsu{3}\times\grpsu[S]{2}$ irreducible tensor operators $\mathcal{O}^{(\lambda_0,\mu_0)S_0}$ between the symplectic many-body basis states to RMEs of the same family of operators between a ket with number of HO quanta decreased by two and a bra with number of HO quanta decreased by two or unchanged. This relation, along with a conjugation relation between RMEs with interchanged ket and bra, can be used to calculate all the RMEs recursively, starting from the RMEs bewteen the LGIs.

The first steps in the derivation of the recurrence relation are to peel the symplectic raising operator $A'^{(2,0)}$ off the ket, as described in Sec.~\ref{part1:subsection:peelA} of Ref.~\cite{part1}, and to commute the raising operator through the operator $\mathcal{O}^{(\lambda_0,\mu_0)S_0}$, as described in Sec.~\ref{part1:subsection:commuting} of Ref.~\cite{part1}. This yields~(\ref{part1:eq:unit_tensor_mb_rme1_generic}) of Ref.~\cite{part1}, which reads
\begin{align}\label{77}
\nonumber\langle\sigma'&\upsilon'\omega'S'\|\mathcal{O}^{(\lambda_0,\mu_0)S_0}\|\sigma\upsilon\omega S\rangle_{\rho_0}\\
\nonumber=&\sum_{\rho_0'}\Phi_{\rho_0\rho_0'}[\omega(\lambda_0,\mu_0)\omega']\sum_{\omega_1\upsilon_1}\sum_{n\rho n_1\rho_1}(K^{\omega}_{\sigma})^{-1}_{\upsilon[n\rho]}\frac{2}{N_n}(n\|b^{\dagger(2,0)}\|n_1)(K^{\omega_1}_{\sigma})_{[n_1\rho_1]\upsilon_1}U[(2,0)n_1\omega\sigma;n_{-}\rho\omega_1\rho_{1-}]\\
\nonumber&\times\sum_{(\lambda_0',\mu_0')\rho_0''}U[(\lambda_0,\mu_0)(2,0)\omega'\omega_1(\lambda_0',\mu_0')_{-}\rho_0''\omega_{-}\rho_0']\sum_{\rho_{00}}\Phi_{\rho_0''\rho_{00}}[(\lambda_0',\mu_0')\omega_1\omega']\\
\nonumber&\times\Big\{\langle\sigma'\upsilon'\omega'S'\|\left[A'^{(2,0)}\times \mathcal{O}^{(\lambda_0,\mu_0)S_0}\right]^{(\lambda_0',\mu_0')S_0}\|\sigma\upsilon_1\omega_1S\rangle_{\rho_{00}}\\
&+\langle\sigma'\upsilon'\omega'S'\|\left[\mathcal{O}^{(\lambda_0,\mu_0)S_0},A'^{(2,0)}\right]^{(\lambda_0',\mu_0')S_0}\|\sigma\upsilon_1\omega_1S\rangle_{\rho_{00}}\Big\},
\end{align}
where the number of HO quanta in the ket is decreased by two, \textit{i.e.}, $N_{\omega_1}=N_{\omega}-2$, $\Phi[\ldots]$ are phase matrices associated with interchanging the coupling order of two $\grpsu{3}$ irreps~\cite{escher1997,escher1998}, $(n\|b^{\dagger(2,0)}\|n_1)$ is an $\grpsu{3}$ RME of the generator $b^{\dagger(2,0)}$ of the u(3)-boson algebra~\cite{rosensteel1981,rowe1982}, $U[\ldots]$ are $\grpsu{3}$ recoupling coefficients\footnote{As a notational convenience, we will often use $\grpu{3}$ labels $\omega$ in expressions for the phase matrices and $\grpsu{3}$ recoupling coefficients, even though only the $\grpsu{3}$ labels $(\lambda_\omega,\mu_\omega)$ are relevant for evaluating these quantities.}~\cite{hecht1965}, and in the last term we have an RME of a coupled commutator of $\mathcal{O}^{(\lambda_0,\mu_0)S_0}$ and $A'^{(2,0)}$ (coupled commutators are reviewed in Appendix~\ref{part1:sec:app-commutator} of Ref.~\cite{part1}). Note that the labels $\gamma$ and $\Sigma$ remain unchanged through the recurrence and will therefore be suppressed in the following.  The $\grpsu{3}$ recoupling coefficients and phase matrices appearing in~(\ref{77}) can be computed using existing codes~\cite{akiyama1973,dytrych2021,ndsu3lib}, while an analytic formula is available for the RME of $b^{\dagger(2,0)}$~\cite{rowe19842,rowe2016,rosensteel1983}, given by~(\ref{part1:n_boson}) of Ref.~\cite{part1}.

The first RME on the right-hand side of~(\ref{77}) can be evaluated using Racah's reduction formula for an RME of a coupled product of two irreducible tensor operators, given by~(B.16) of Ref.~\cite{mccoy2018dis}. The result is
\begin{align}\label{racahred}
\nonumber\langle\sigma'\upsilon'\omega'S'\|&\left[A'^{(2,0)}\times \mathcal{O}^{(\lambda_0,\mu_0)S_0}\right]^{(\lambda_0',\mu_0')S_0}\|\sigma\upsilon_1\omega_1;S\rangle_{\rho_{00}}\\
\nonumber=&\sum_{\upsilon''\omega''\rho''}(-1)^{\lambda_0+\mu_0-\lambda_0'-\mu_0'}U[\omega_1(\lambda_0,\mu_0)\omega'(2,0);\omega''\rho''_{-}(\lambda_0',\mu_0')_{-}\rho_{00}]\\
&\times\langle\sigma'\upsilon'\omega'S'\|A'^{(2,0)}\|\sigma'\upsilon''\omega''S'\rangle\langle\sigma'\upsilon''\omega''S'\|\mathcal{O}^{(\lambda_0,\mu_0)S_0}\|\sigma\upsilon_1\omega_1S\rangle_{\rho''},
\end{align}
where we used the fact that the symplectic raising operator does not act on the spin degrees of freedom and connects only states belonging to the same symplectic irrep, and with numbers of HO quanta differing by two, which means that $N_{\omega''}=N_{\omega'}-2$. Analytic formulae for the RMEs of the symplectic raising operator between the symplectic basis states appearing in~(\ref{racahred}) are readily available from VCS theory~\cite{rowe1984,rowe19842,hecht1987,rowe1995,rowe2016}\footnote{VCS theory relates the symplectic generators to the generators of a simpler u(3)-boson algebra. This allows for the evaluation of matrix elements of the symplectic generators by relating them to the known matrix elements of the generators of the u(3)-boson algebra.} and given in Secs.~\ref{part1:subsection:sp3r_irrep} and~\ref{part1:subsection:sp3r-generator-rme} of~\cite{part1}.

The next step is to express the coupled commutator appearing in the last term of~(\ref{77}) in terms of the family of operators $\mathcal{O}^{(\lambda_0,\mu_0)S_0}$. The intrinsic symplectic raising operator $A'^{(2,0)}$ used to construct the SpNCCI basis consists of a one-body part and a two-body part. This would seem to be a problem, because the commutator of an $n$-body operator $\mathcal{O}^{(\lambda_0,\mu_0)S_0}$ and a two-body operator in general contains an ($n+1$)-body part.  If the operator $\mathcal{O}^{(\lambda_0,\mu_0)S_0}$ is a one-body operator, the two-body part of the symplectic raising operator would be expected to induce a two-body term, \textit{etc.}  In general, the particle rank of the operators under consideration would be expected to explode with successive application of the relation~(\ref{77}), which would therefore not in fact provide a useable recurrence relation.

However, for the case in which the operator $\mathcal{O}^{(\lambda_0,\mu_0)S_0}$ is intrinsic~\cite{reske1984,suzuki1986,part1}, no ($n+1$)-body part of the commutator is induced, as shown in Sec.~\ref{part1:subsection:evaluate:AUcom} and Appendix~\ref{part1:sec:app-coupled-commutator-unit-tensors} of Ref.~\cite{part1}.  Thus, for matrix elements of \textit{intrinsic} two-body operators, a recurrence relation, (\ref{part1:eq:recurrence_final}) of Ref.~\cite{part1}, can indeed be obtained.

The same argument cannot be applied in the cases of present interest, namely, for general $n$-body density matrix elements.  These may be viewed as matrix elements of fundamental $n$-body operators, which in general depend on the single-particle coordinates.  Nonetheless, we shall see in the following that recurrence relations can still be derived, through careful consideration of the center-of-mass correction to the symplectic raising operator and its action on a basis which is free of center-of-mass excitations.

\section{Recurrence for density matrix elements between symplectic many-body basis states}
\label{4.recu}

We now describe the procedure to recursively calculate the density matrix elements, reduced with respect to $\grpsu{3}\times\grpsu[S]{2}$, between the symplectic many-body basis states, from the RMEs between the LGIs as the seeds for the recurrence. The derivation of the present recurrence contains important differences from the case of that for the RMEs of an intrinsic operator, described in Sec.~\ref{part1:section:recurrence} of Ref.~\cite{part1}. We present results for the one-body and two-body density RMEs (OBDRMEs and TBDRMEs), which are of the greatest interest in many-body calculations. In principle, the procedure can be applied to densities of arbitrary particle rank (thereby allowing one to calculate matrix elements of an arbitrary operator).  However, for densities of higher particle rank the results would be progressively more complex.

We start the derivation from~(\ref{77}), where now, to obtain OBDRMEs we take the operator $\mathcal{O}^{(\lambda_0,\mu_0)S_0}$ to be a fundamental one-body operator, 
\begin{equation}\label{obd}
\mathcal{O}^{(\lambda_0,\mu_0)S_0}=\left[a^{\dagger}_{N'}\times\tilde{a}_{N}\right]^{(\lambda_0,\mu_0)S_{0}},
\end{equation}
or, for the TBDRMEs, a fundamental two-body operator
\begin{equation}\label{tbd}
\mathcal{O}^{(\lambda_0,\mu_0)S_0}=\left[\left[a^{\dagger}_{N_1}\times a^{\dagger}_{N_2}\right]^{(\lambda_f,\mu_f)S_{f}}\times\left[\tilde{a}_{N_3}\times\tilde{a}_{N_4}\right]^{(\lambda_i,\mu_i)S_{i}}\right]^{\rho(\lambda_0,\mu_0)S_{0}},
\end{equation}
where the label $\rho$ distinguishes different $\grpsu{3}$ irreps with the same labels $(\lambda_0,\mu_0)$ resulting from the coupling of $(\lambda_f,\mu_f)$ and $(\lambda_i,\mu_i)$. Now we need to express the two RMEs on the right-hand side of~(\ref{77}) in terms of OBDRMEs, if $\mathcal{O}^{(\lambda_0,\mu_0)S_0}$ is given by~(\ref{obd}), or TBDRMEs, if $\mathcal{O}^{(\lambda_0,\mu_0)S_0}$ is given by~(\ref{tbd}). For the first RME we can use the result~(\ref{racahred}) of applying Racah’s reduction formula. In the second RME we express the intrinsic symplectic raising operator $A'^{(2,0)}$ in terms of the laboratory-frame symplectic raising operator $A^{(2,0)}$ and the center-of-mass correction $A^{(2,0)}_{\rm cm}$, as in~(\ref{syraop}):
\begin{align}\label{der5}
\nonumber\langle\sigma'\upsilon'\omega'S'&\|\left[\mathcal{O}^{(\lambda_0,\mu_0)S_0},A'^{(2,0)}\right]^{(\lambda_0',\mu_0')S_0}\|\sigma\upsilon_1\omega_1S\rangle_{\rho_{00}}\\
\nonumber=&\langle\sigma'\upsilon'\omega'S'\|\left[\mathcal{O}^{(\lambda_0,\mu_0)S_0},A^{(2,0)}\right]^{(\lambda_0',\mu_0')S_0}\|\sigma\upsilon_1\omega_1S\rangle_{\rho_{00}}\\
&-\langle\sigma'\upsilon'\omega'S'\|\left[\mathcal{O}^{(\lambda_0,\mu_0)S_0},A^{(2,0)}_{\rm cm}\right]^{(\lambda_0',\mu_0')S_0}\|\sigma\upsilon_1\omega_1S\rangle_{\rho_{00}}.
\end{align}
Recall that $A^{(2,0)}$ is a one-body operator and $A^{(2,0)}_{\rm cm}$, which needs to be subtracted from $A^{(2,0)}$, contains both one-body and two-body parts.

It is straightforward to express the first RME on the right-hand side of~(\ref{der5}) in terms of density RMEs. Since $A^{(2,0)}$ is a one-body operator, its commutator with an $n$-body operator $\mathcal{O}^{(\lambda_0,\mu_0)S_0}$ is still an $n$-body operator, and no higher-body terms are induced.  In particular, one simply needs to evaluate the coupled commutator of the operator $\mathcal{O}^{(\lambda_0,\mu_0)S_0}$ and the laboratory-frame symplectic raising operator $A^{(2,0)}$, with the latter expressed in terms of the fermion creation and annihilation operators as in~(\ref{4.6}).\footnote{For evaluation of the commutator, the formulae (B8) and (B11) in Ref.~\cite{escher1998} for commutators of coupled products of the fermion creation and annihilation operators are useful.}

However, the center-of-mass correction to the symplectic raising operator appearing in the coupled commutator in the second RME on the right-hand side of~(\ref{der5}) contains a two-body part, and a commutator of a two-body operator and an $n$-body operator $\mathcal{O}^{(\lambda_0,\mu_0)S_0}$ is, in general, an ($n+1$)-body operator. If $\mathcal{O}^{(\lambda_0,\mu_0)S_0}$ is an intrinsic operator (see Sec.~\ref{part1:section:recurrence} of Ref.~\cite{part1}), no higher-body part of the commutator is induced~\cite{reske1984,suzuki1986,part1}. But this is not true for the operators~(\ref{obd}) and~(\ref{tbd}). Nevertheless, it is still possible to derive a recurrence relation for the OBDRMEs and TBDRMEs (or matrix elements of an arbitrary $n$-body operator) between the symplectic many-body basis states using the fact, as suggested by Reske~\cite{reske1984}, that the symplectic many-body basis states are free of center-of-mass excitations.  To exploit this property, we express the center-of-mass correction to the symplectic raising operator in the form~(\ref{4.8}), that is, explicitly in terms of the one-body operator $c^{\dagger(1,0)}_{\rm cm}$ creating an HO quantum in the center-of-mass motion.

The idea is as follows.  The coupled commutator appearing in the second RME on the right-hand side of~(\ref{der5}), that is, of the operator $\mathcal{O}^{(\lambda_0,\mu_0)S_0}$ and the center-of-mass correction~(\ref{4.8}) to the symplectic raising operator, can be written as
\begin{equation}\label{der7}
\left[\mathcal{O}^{(\lambda_0,\mu_0)S_0},A^{(2,0)}_{\rm cm}\right]^{(\lambda_0',\mu_0')S_0}=\frac{1}{\sqrt{2}}\left[\mathcal{O}^{(\lambda_0,\mu_0)S_0},\left[c^{\dagger(1,0)}_{\rm cm}\times c^{\dagger(1,0)}_{\rm cm}\right]^{(2,0)}\right]^{(\lambda_0',\mu_0')S_0}.
\end{equation}
To evaluate the coupled commutator in~(\ref{der7}), we use the identity~(\ref{part1:eqn:coupled-commutator-su3-product}) of Ref.~\cite{part1}, which is the analog for coupled commutators of the well-known commutator identity $[A,BC]=[A,B]C+B[A,C]$. We obtain
\begin{align}\label{der8}
\nonumber\left[\mathcal{O}^{(\lambda_0,\mu_0)S_0},A^{(2,0)}_{\rm cm}\right]^{(\lambda_0',\mu_0')S_0}&=\frac{1}{\sqrt{2}}\sum_{(\lambda,\mu)}U[(\lambda_0,\mu_0)(1,0)(\lambda_0',\mu_0')(1,0);(\lambda,\mu)_{--}(2,0)_{--}]\\
\nonumber&\times\Bigg(\left[\left[\mathcal{O}^{(\lambda_0,\mu_0)S_0},c^{\dagger(1,0)}_{\rm cm}\right]^{(\lambda,\mu)S_0}\times c^{\dagger(1,0)}_{\rm cm}\right]^{(\lambda_0',\mu_0')S_0}\\
&-(-1)^{\lambda+\mu-\lambda_0'-\mu_0'}\left[c^{\dagger(1,0)}_{\rm cm}\times\left[\mathcal{O}^{(\lambda_0,\mu_0)S_0},c^{\dagger(1,0)}_{\rm cm}\right]^{(\lambda,\mu)S_0}\right]^{(\lambda_0',\mu_0')S_0}\Bigg).
\end{align}
Using the definition of the coupled commutator given by~(\ref{part1:eqn:coupled-commutator-su3-su2-difference}) of Ref.~\cite{part1}, the expression~(\ref{der8}) can be further rewritten as
\begin{align}\label{der9}
\nonumber\left[\mathcal{O}^{(\lambda_0,\mu_0)S_0},A^{(2,0)}_{\rm cm}\right]^{(\lambda_0',\mu_0')S_0}&=\frac{1}{\sqrt{2}}\sum_{(\lambda,\mu)}U[(\lambda_0,\mu_0)(1,0)(\lambda_0',\mu_0')(1,0);(\lambda,\mu)_{--}(2,0)_{--}]\\
\nonumber&\times\Bigg(\left[\left[\mathcal{O}^{(\lambda_0,\mu_0)S_0},c^{\dagger(1,0)}_{\rm cm}\right]^{(\lambda,\mu)S_0},c^{\dagger(1,0)}_{\rm cm}\right]^{(\lambda_0',\mu_0')S_0}\\
&-2(-1)^{\lambda+\mu-\lambda_0'-\mu_0'}\left[c^{\dagger(1,0)}_{\rm cm}\times\left[\mathcal{O}^{(\lambda_0,\mu_0)S_0},c^{\dagger(1,0)}_{\rm cm}\right]^{(\lambda,\mu)S_0}\right]^{(\lambda_0',\mu_0')S_0}\Bigg).
\end{align}
However, since the operator $c^{\dagger(1,0)}_{\rm cm}$ creates an HO quantum in the center-of-mass motion, and the symplectic many-body basis states are free of HO excitations in the center-of-mass motion, the second term on the right hand side of~(\ref{der9}) annihilates the bra, and thus does not contribute to the second RME on the right-hand side of~(\ref{der5}).  Only the first term in~(\ref{der9}) can contribute to the RME.  If $\mathcal{O}^{(\lambda_0,\mu_0)S_0}$ is an $n$-body operator, this term is still an $n$-body operator (recall that $c^{\dagger(1,0)}_{\rm cm}$ is a one-body operator, and a commutator of an $n$-body and a one-body operator is still an $n$-body operator). Therefore, no ($n+1$)-body terms are induced in the coupled commutators on the right-hand side of~(\ref{der5}). This allows for derivation of a recurrence relation for the $n$-body density RMEs (or matrix elements of the fundamental $n$-body operators).

To express the second RME on the right-hand side of~(\ref{der5}) in terms of density RMEs, one needs to evaluate the nested coupled commutator in the first term in~(\ref{der9}), where the center-of-mass HO raising operator $c^{\dagger(1,0)}_{\rm cm}$ is given by~(\ref{Bdag}). For $\mathcal{O}^{(\lambda_0,\mu_0)S_0}$ given by~(\ref{obd}) or~(\ref{tbd}) this can be done using~(B8) and~(B11) of Ref.~\cite{escher1998} for commutators of coupled products of the fermion creation and annihilation operators. After thereby reexpressing the RME and inserting it into the second term on the right hand side of~(\ref{77}), one obtains the recurrence relations for the density RMEs between the symplectic many-body basis states.
  
For the OBDRMEs the recurrence relation reads\footnote{To obtain the result~(\ref{recurrence1}) one also needs to use relations between $\grpsu{3}$ recoupling and coupling coefficients and symmetry and orthonormality relations for the coupling coefficients, which can be found in Appendix C of Ref.~\cite{escher1998}, as well as symmetry properties and special values of $\grpsu{3}$ recoupling coefficients, which can be found in Appendix A of Ref.~\cite{hecht1981}.}
\begin{align}\label{recurrence1}
\nonumber\langle&\sigma'\upsilon'\omega'S'\|\left[a^{\dagger}_{N'}\times\tilde{a}_{N}\right]^{(\lambda_0,\mu_0)S_{0}}\|\sigma\upsilon\omega S\rangle_{\rho_{0}}=\sum_{\omega_{1}\upsilon_{1}}\sum_{n\rho n_1\rho_1}(K^{\omega}_{\sigma})^{-1}_{\upsilon[n\rho]}\frac{2}{N_n}(n\|b^{\dagger(2,0)}\|n_1)\\
\nonumber&\times U[(2,0)n_1\omega\sigma;n_{-}\rho\omega_1\rho_{1-}](K^{\omega_1}_{\sigma})_{[n_1\rho_1]\upsilon_1}\sqrt{\frac{\dim\omega}{\dim(\lambda_0,\mu_0)\dim\omega_{1}}}\\
&\times\Bigg\{A_{0,0}+\sum_{(\lambda_0',\mu_0')\rho'}U[\omega'(\mu_0',\lambda_0')\omega(2,0);\omega_{1}\rho'_{-}(\mu_0,\lambda_0)_{-}\rho_{0}]\sqrt{\dim(\lambda_0',\mu_0')}\big(A_{0,2}-A_{2,0}-A_{1,1}\big)\Bigg\},
\end{align}
where $\dim(\lambda,\mu)=(\lambda+1)(\mu+1)(\lambda+\mu+2)/2$ is the dimension of the $\grpsu{3}$ irrep.  The terms denoted by $A$, with various subscripts, are defined below.  They again involve OBDRMEs, but with fewer HO quanta in the states and, in general, different HO shell indices on the creation and annihilation operators appearing in the OBDRME, where the subscripts serve to indicate these differences. The recurrence~(\ref{recurrence1}) is thus, equivalently, a recurrence relation for RMEs of fundamental one-body operators in terms of RMEs of other fundamental one-body operators, at lower number of HO quanta for the bra and/or ket states.

The term $A_{0,0}$ contains RMEs of the same fundamental one-body operator as appears on the left-hand side of~(\ref{recurrence1}), that is, with unchanged values of the HO shell indices $N'$ and $N$ appearing on the creation and annihilation operators, but now between states in which the number of HO quanta, for both the bra and ket, is reduced by two relative to the states appearing on the left-hand side of~(\ref{recurrence1}):
\begin{align}
\nonumber A_{0,0}=&(-1)^{\lambda_{0}+\mu_0+\lambda_{\omega}+\mu_{\omega}+\lambda_{\omega}'+\mu_{\omega}'}\sum_{\omega''\upsilon''\rho'}\sqrt{\frac{\dim\omega_{1}\dim\omega''}{6}}\\
\nonumber&\times\sum_{\rho_{2}\rho_{3}}\Phi_{\rho'\rho_{2}}[\omega'',(\mu_0,\lambda_0);\omega_{1}]\Phi_{\rho_{2}\rho_{3}}[\omega_{1},\tilde{\omega}'';(\mu_0,\lambda_0)]U[\omega'\tilde{\omega}''\omega\omega_{1};(2,0)_{--}(\mu_0,\lambda_0)\rho_{3}\rho_{0}]\\
&\times\langle\sigma'\upsilon'\omega'S'\|A'^{(20)}\|\sigma'\upsilon''\omega''S'\rangle\langle\sigma'\upsilon''\omega''S'\|\left[a^{\dagger}_{N'}\times\tilde{a}_{N}\right]^{(\lambda_0,\mu_0)S_{0}}\|\sigma\upsilon_{1}\omega_{1}S\rangle_{\rho'},
\end{align}
where $N_{\omega''}=N_{\omega'}-2$, $N_{\omega_1}=N_{\omega}-2$, and a tilde on an $\grpsu{3}$ irrep label denotes conjugation, \textit{i.e.}, $(\lambda,\mu) \leftrightarrow (\mu,\lambda)$. The remaining terms in~(\ref{recurrence1}) contain OBDRMEs in which the bra state is the same as appears on the left-hand side of~(\ref{recurrence1}), while the number of HO quanta in the ket state is reduced by two. The fundamental one-body operator appearing in the OBDRME in these terms is no longer that appearing on the left-hand side of~(\ref{recurrence1}), but rather differs in the values of one or both of the HO shell indices, such that the net number of HO quanta created by this operator is larger, by two, than the number created by the original operator. In particular, the term $A_{0,2}$ contains OBDRMEs with $N$ reduced by two compared to the OBDRME on the left-hand side of~(\ref{recurrence1}):
\begin{align}
\nonumber A_{0,2}=&\left(1-\frac{1}{A}\right)\sqrt{\dim(N,0)}U[(N',0)(0,N)(\lambda_0',\mu_0')(2,0);(\lambda_0,\mu_0)_{--}(0,N-2)_{--}]\\
&\times\langle\sigma'\upsilon'\omega'S'\|\left[a^{\dagger}_{N'}\times\tilde{a}_{N-2}\right]^{(\lambda_0',\mu_0')S_{0}}\|\sigma\upsilon_{1}\omega_{1}S\rangle_{\rho'}.
\end{align}
The term $A_{2,0}$ contains OBDRMEs with $N'$ increased by two:
\begin{align}\label{recurrence1-a20}
\nonumber A_{2,0}=&\left(1+\frac{1}{A}\right)(-1)^{\lambda_{0}+\mu_0+\lambda_{0}'+\mu_0'}\sqrt{\dim(N',0)}\\
\nonumber&\times U[(2,0)(N',0)(\lambda_0',\mu_0')(0,N);(N'+2,0)_{--}(\lambda_0,\mu_0)_{--}]\\
&\times\langle\sigma'\upsilon'\omega'S'\|\left[a^{\dagger}_{N'+2}\times\tilde{a}_{N}\right]^{(\lambda_0',\mu_0')S_{0}}\|\sigma\upsilon_{1}\omega_{1}S\rangle_{\rho'}.
\end{align}
Finally, the term $A_{1,1}$ contains OBDRMEs with $N'$ increased by one and $N$ reduced by one:
\begin{align}
\nonumber A_{1,1}=&\frac{1}{A}\sqrt{2(N'+1)(N+2)}\sum_{(\lambda,\mu)}(-1)^{\lambda_{0}'+\mu_0'+\lambda+\mu}\\
\nonumber&\times U[(N',0)(0,N)(\lambda,\mu)(1,0);(\lambda_0,\mu_0)_{--}(0,N-1)_{--}]\\
\nonumber&\times U[(1,0)(N',0)(\lambda_0',\mu_0')(0,N-1);(N'+1,0)_{--}(\lambda,\mu)_{--}]\\
\nonumber&\times U[(\lambda_0,\mu_0)(1,0)(\lambda_0',\mu_0')(1,0);(\lambda,\mu)_{--}(2,0)_{--}]\\
&\times\langle\sigma'\upsilon'\omega'S'\|\left[a^{\dagger}_{N'+1}\times\tilde{a}_{N-1}\right]^{(\lambda_0',\mu_0')S_{0}}\|\sigma\upsilon_{1}\omega_{1}S\rangle_{\rho'}.
\end{align}
The equation~(\ref{recurrence1}) relates the OBDRMEs between the symplectic many-body basis states with OBDRMEs between basis states with numbers of HO quanta reduced by two. The recurrence relation~(\ref{recurrence1}), along with the conjugation relation
\begin{align}\label{con1}
\nonumber\langle\sigma'\upsilon'\omega'S'\|\left[a^{\dagger}_{N'}\times\tilde{a}_{N}\right]^{(\lambda_0,\mu_0)S_0}\|\sigma\upsilon\omega S\rangle_{\rho_0}=(-1)^{N'-N+\lambda_{\omega}'+\mu_{\omega}'-\lambda_{\omega}-\mu_{\omega}+S'-S}\\
\times\sqrt{\frac{(2S+1)\dim\omega}{(2S'+1)\dim\omega'}}\langle\sigma\upsilon\omega S\|\left[a^{\dagger}_{N}\times\tilde{a}_{N'}\right]^{(\mu_0,\lambda_0)S_0}\|\sigma'\upsilon'\omega'S'\rangle_{\rho_0}^{*},
\end{align}
can be used to calculate the OBDRMEs recursively from the OBDRMEs between the LGIs.

The recurrence relation~(\ref{recurrence1}) for the OBDRMEs between the symplectic many-body basis states is similar to the recurrence relation derived in Refs.~\cite{escher1997,escher1998}, but the contribution of the two-body part of the symplectic raising operator was incorrectly assumed to vanish in that work. Compared to that work, the recurrence relation~(\ref{recurrence1}) contains an additional term $A_{1,1}$, while the $1/A$ term appears with opposite sign in the prefactor in the expression~(\ref{recurrence1-a20}) for $A_{2,0}$.

For the TBDRMEs the recurrence relation reads
\begin{align}\label{recurrence2}
\nonumber\langle\sigma'\upsilon'&\omega'S'\|\left[\left[a_{N_1}^{\dagger}\times a_{N_2}^{\dagger}\right]^{(\lambda_f,\mu_f)S_f}\times\left[\tilde{a}_{N_3}\times\tilde{a}_{N_4}\right]^{(\lambda_i,\mu_i)S_{i}}\right]^{\rho_0(\lambda_0,\mu_0)S_0}\|\sigma\upsilon\omega S\rangle_{\bar{\rho}}\\
\nonumber=&(-1)^{\lambda_0+\mu_0}\sum_{\upsilon_{1}\omega_{1}}\sum_{n\rho n_1\rho_1}(K^{\omega}_{\sigma})^{-1}_{\upsilon[n\rho]}\frac{2}{N_n}(n\|b^{\dagger(2,0)}\|n_1)U[(2,0)n_1\omega\sigma;n_{-}\rho\omega_1\rho_{1-}](K^{\omega_1}_{\sigma})_{[n_1\rho_1]\upsilon_1}\\
\nonumber&\times\Bigg\{A_{0,0,0,0}+\sum_{(\lambda_0',\mu_0')\rho_0'\rho'\rho_2\rho_3}\Phi_{\rho'\rho_2}[\omega_{1},(\lambda_0',\mu_0');\omega']\Phi_{\rho_3\bar{\rho}}[(\lambda_0,\mu_0),\omega;\omega']\\
\nonumber&\times U[(\lambda_0,\mu_0)(2,0)\omega'\omega_{1};(\lambda_0',\mu_0')_{-}\rho_2\omega_{-}\rho_3]\big(A_{0,0,0,2}+A_{0,0,2,0}+A_{2,0,0,0}+A_{0,2,0,0}+A_{0,0,1,1}\\
&+A_{1,0,0,1}+A_{0,1,0,1}+A_{1,0,1,0}+A_{0,1,1,0}+A_{1,1,0,0}\big)\Bigg\},
\end{align}
where the terms denoted by $A$, with various subscripts, involve RMEs of fundamental two-body operators with HO shell indices on the creation and annihilation operators changed compared to those on the left-hand side by values indicated by the subscripts. The term $A_{0,0,0,0}$ contains RMEs of the same fundamental two-body operator as appears on the left-hand side of~(\ref{recurrence2}), that is, with unchanged values of the HO shell indices $N_1$, $N_2$, $N_3$, and $N_4$ appearing on the creation and annihilation operators, but now between states in which the number of HO quanta, for both the bra and ket, is reduced by two relative to the states appearing on the left-hand side of~(\ref{recurrence2}):
\begin{align}
\nonumber A_{0,0,0,0}=&\sum_{\upsilon''\omega''\rho'}(-1)^{\lambda_{\omega}+\mu_{\omega}+\lambda_{\omega'}+\mu_{\omega'}}\sqrt{\frac{\dim\omega\dim\omega''}{6\dim(\lambda_0,\mu_0)}}\sum_{\rho_{2}\rho_{3}}\Phi_{\rho'\rho_{2}}[\omega'',(\mu_0,\lambda_0);\omega_{1}]\Phi_{\rho_{2}\rho_{3}}[\omega_{1},\tilde{\omega}'';(\mu_0,\lambda_0)]\\
\nonumber&\times U[\omega'\tilde{\omega}''\omega\omega_{1};(2,0)_{--}(\mu_0,\lambda_0)\rho_{3}\bar{\rho}]\langle\sigma'\upsilon'\omega'S'\|A'^{(2,0)}\|\sigma'\upsilon''\omega''S'\rangle\\
&\times\langle\sigma'\upsilon''\omega''S'\|\left[\left[a_{N_1}^{\dagger}\times a_{N_2}^{\dagger}\right]^{(\lambda_f,\mu_f)S_{f}}\times\left[\tilde{a}_{N_3}\times\tilde{a}_{N_4}\right]^{(\lambda_i,\mu_i)S_{i}}\right]^{\rho_0(\lambda_0,\mu_0)S_0}\|\sigma\upsilon_{1}\omega_{1}S\rangle_{\rho'},
\end{align}
where $N_{\omega''}=N_{\omega'}-2$ and $N_{\omega_1}=N_{\omega}-2$. The remaining terms contain TBDRMEs in which the bra state is the same as appears on the left-hand side of~(\ref{recurrence2}), while the number of HO quanta in the ket state is reduced by two.  The fundamental two-body operator appearing in the TBDRME in these terms is no longer that appearing on the left-hand side of~(\ref{recurrence2}), but rather differs in the values of one or two of the HO shell indices, such that the net number of HO quanta created by the operator is increased by two.  In particular, the term $A_{0,0,0,2}$ contains TBDRMEs with $N_4$ reduced by two compared to the TBDRME on the left-hand side of~(\ref{recurrence2}):
\begin{align}
\nonumber A_{0,0,0,2}&=\frac{(-1)^{\lambda_{f}+\mu_f-\lambda_{i}-\mu_i}}{6}\sqrt{\frac{\dim(\lambda_0,\mu_0)\dim(\lambda_i,\mu_i)(N_4-1)N_4(N_4+1)(N_4+2)}{\dim(\lambda_f,\mu_f)\dim(0,N_3)}}\\
\nonumber&\times\sum_{(\lambda,\mu)}U[(\lambda_0,\mu_0)(\mu_i,\lambda_i)(\lambda_0',\mu_0')(\lambda,\mu);(\lambda_f,\mu_f)\rho_0\rho_{0}'(2,0)_{--}]\\
\nonumber&\times\bigg\{\frac{(-1)^{\lambda+\mu+N_3+N_4}}{2}\sqrt{\dim(\lambda,\mu)}U[(\lambda_i,\mu_i)(N_4,0)(\lambda,\mu)(0,N_4-2);(0,N_3)_{--}(2,0)_{--}]\\
\nonumber&-\frac{1}{A}\sqrt{\frac{N_4(N_4+1)}{6\dim(0,N_3)}}\sum_{(\lambda',\mu')}\dim(\lambda',\mu')U[(\lambda_i,\mu_i)(N_4,0)(\lambda',\mu')(0,N_4-1);(0,N_3)_{--}(1,0)_{--}]\\
\nonumber&\times U[(\mu_i,\lambda_i)(\lambda',\mu')(2,0)(1,0);(1,0)_{--}(\lambda,\mu)_{--}]\\
\nonumber&\times U[(\lambda',\mu')(N_4-1,0)(\lambda,\mu)(0,N_4-2);(0,N_3)_{--}(1,0)_{--}]\bigg\}\\
&\times\langle\sigma'\upsilon'\omega'S'\|\left[\left[a_{N_1}^{\dagger}\times a_{N_2}^{\dagger}\right]^{(\lambda_f,\mu_f)S_{f}}\times\left[\tilde{a}_{N_3}\times\tilde{a}_{N_4-2}\right]^{(\lambda,\mu) S_{i}}\right]^{\rho_{0}'(\lambda_0',\mu_0')S_0}\|\sigma\upsilon_{1}\omega_{1}S\rangle_{\rho'}.
\end{align}
The term $A_{0,0,2,0}$ contains TBDRMEs with $N_3$ reduced by two:
\begin{align}
\nonumber&A_{0,0,2,0}=\frac{(-1)^{\lambda_{f}+\mu_f}}{6}\sqrt{\frac{\dim(\lambda_0,\mu_0)\dim(\lambda_i,\mu_i)(N_3-1)N_3(N_3+1)(N_3+2)}{\dim(\lambda_f,\mu_f)\dim(0,N_4)}}\\
\nonumber&\times\sum_{(\lambda,\mu)}U[(\lambda_0,\mu_0)(\mu_i,\lambda_i)(\lambda_0',\mu_0')(\lambda,\mu);(\lambda_f,\mu_f)\rho_0\rho_{0}'(2,0)_{--}]\\
\nonumber&\times\bigg\{\frac{(-1)^{N_3+N_4}}{2}\sqrt{\dim(\lambda,\mu)}U[(\lambda_i,\mu_i)(N_3,0)(\lambda,\mu)(0,N_3-2);(0,N_4)_{--}(2,0)_{--}]\\
\nonumber&-\frac{(-1)^{\lambda+\mu}}{A}\sqrt{\frac{N_3(N_3+1)}{6\dim(0,N_4)}}\sum_{(\lambda',\mu')}\dim(\lambda',\mu')U[(\lambda_i,\mu_i)(N_3,0)(\lambda',\mu')(0,N_3-1);(0,N_4)_{--}(1,0)_{--}]\\
\nonumber&\times U[(\mu_i,\lambda_i)(\lambda',\mu')(2,0)(1,0);(1,0)_{--}(\lambda,\mu)_{--}]\\
\nonumber&\times U[(\lambda',\mu')(N_3-1,0)(\lambda,\mu)(0,N_3-2);(0,N_4)_{--}(1,0)_{--}]\bigg\}\\
&\times\langle\sigma'\upsilon'\omega'S'\|\left[\left[a_{N_1}^{\dagger}\times a_{N_2}^{\dagger}\right]^{(\lambda_f,\mu_f)S_{f}}\times\left[\tilde{a}_{N_3-2}\times\tilde{a}_{N_4}\right]^{(\lambda,\mu) S_{i}}\right]^{\rho_{0}'(\lambda_0',\mu_0')S_0}\|\sigma\upsilon_{1}\omega_{1}S\rangle_{\rho'}.
\end{align}
The term $A_{2,0,0,0}$ contains TBDRMEs with $N_1$ increased by two:
\begin{align}
\nonumber A_{2,0,0,0}=&-(-1)^{\lambda_{0}'+\mu_0'}\sqrt{\frac{(N_1+1)(N_1+2)}{2}}\\
\nonumber&\times\sum_{(\lambda,\mu)}U[(2,0)(\lambda_f,\mu_f)(\lambda_0',\mu_0')(\lambda_i,\mu_i);(\lambda,\mu)_{-}\rho_{0}'(\lambda_0,\mu_0)\rho_{0-}]\\
\nonumber&\times\bigg\{U[(2,0)(N_1,0)(\lambda,\mu)(N_2,0);(N_1+2,0)_{--}(\lambda_f,\mu_f)_{--}]\\
\nonumber&+\frac{1}{A}\sum_{(\lambda',\mu')}U[(1,0)(N_1,0)(\lambda',\mu')(N_2,0);(N_1+1,0)_{--}(\lambda_f,\mu_f)_{--}]\\
\nonumber&\times U[(\lambda_f,\mu_f)(1,0)(\lambda,\mu)(1,0);(\lambda',\mu')_{--}(2,0)_{--}]\\
\nonumber&\times U[(1,0)(N_1+1,0)(\lambda,\mu)(N_2,0);(N_1+2,0)_{--}(\lambda',\mu')_{--}]\bigg\}\\
&\times\langle\sigma'\upsilon'\omega'S'\|\left[\left[a_{N_1+2}^{\dagger}\times a_{N_2}^{\dagger}\right]^{(\lambda,\mu) S_{f}}\times\left[\tilde{a}_{N_3}\times\tilde{a}_{N_4}\right]^{(\lambda_i,\mu_i)S_{i}}\right]^{\rho_{0}'(\lambda_0',\mu_0')S_0}\|\sigma\upsilon_{1}\omega_{1}S\rangle_{\rho'}.
\end{align}
The term $A_{0,2,0,0}$ contains TBDRMEs with $N_2$ increased by two:
\begin{align}
\nonumber A_{0,2,0,0}=&-(-1)^{\lambda_{f}+\mu_f+\lambda_{0}'+\mu_0'}\sqrt{\frac{(N_2+1)(N_2+2)}{2}}\\
\nonumber&\times\sum_{(\lambda,\mu)}(-1)^{\lambda+\mu}U[(2,0)(\lambda_f,\mu_f)(\lambda_0',\mu_0')(\lambda_i,\mu_i);(\lambda,\mu)_{-}\rho_{0}'(\lambda_0,\mu_0)\rho_{0-}]\\
\nonumber&\times\bigg\{U[(2,0)(N_2,0)(\lambda,\mu)(N_1,0);(N_2+2,0)_{--}(\lambda_f,\mu_f)_{--}]\\
\nonumber&+\frac{1}{A}\sum_{(\lambda',\mu')}U[(1,0)(N_2,0)(\lambda',\mu')(N_1,0);(N_2+1,0)_{--}(\lambda_f,\mu_f)_{--}]\\
\nonumber&\times U[(\lambda_f,\mu_f)(1,0)(\lambda,\mu)(1,0);(\lambda',\mu')_{--}(2,0)_{--}]\\
\nonumber&\times U[(1,0)(N_2+1,0)(\lambda,\mu)(N_1,0);(N_2+2,0)_{--}(\lambda',\mu')_{--}]\bigg\}\\
&\times\langle\sigma'\upsilon'\omega'S'\|\left[\left[a_{N_1}^{\dagger}\times a_{N_2+2}^{\dagger}\right]^{(\lambda,\mu) S_{f}}\times\left[\tilde{a}_{N_3}\times\tilde{a}_{N_4}\right]^{(\lambda_i,\mu_i)S_i}\right]^{\rho_{0}'(\lambda_0',\mu_0')S_0}\|\sigma\upsilon_{1}\omega_{1}S\rangle_{\rho'}.
\end{align}
The term $A_{0,0,1,1}$ contains TBDRMEs with $N_3$ and $N_4$ reduced by one:
\begin{align}
\nonumber A_{0,0,1,1}&=\frac{(-1)^{\lambda_{f}+\mu_f}}{3A}\sqrt{\frac{\dim(\lambda_0,\mu_0)\dim(\lambda_i,\mu_i)}{6\dim(\lambda_f,\mu_f)}}\\
\nonumber&\times\sum_{(\lambda,\mu)(\lambda',\mu')}(-1)^{\lambda'+\mu'}\dim(\lambda',\mu')U[(\lambda_0,\mu_0)(\mu_i,\lambda_i)(\lambda_0',\mu_0')(\lambda,\mu);(\lambda_f,\mu_f)\rho_0\rho_{0}'(2,0)_{--}]\\
\nonumber&\times U[(\mu_i,\lambda_i)(\lambda',\mu')(2,0)(1,0);(1,0)_{--}(\lambda,\mu)_{--}]\\
\nonumber&\times\Big\{(-1)^{\lambda_{i}+\mu_i+\lambda+\mu}\sqrt{N_3(N_4+2)}U[(\lambda_i,\mu_i)(N_4,0)(\lambda',\mu')(0,N_4-1);(0,N_3)_{--}(1,0)_{--}]\\
\nonumber&\times U[(\lambda',\mu')(N_3,0)(\lambda,\mu)(0,N_3-1);(0,N_4-1)_{--}(1,0)_{--}]\\
\nonumber&+\sqrt{N_4(N_3+2)}U[(\lambda_i,\mu_i)(N_3,0)(\lambda',\mu')(0,N_3-1);(0,N_4)_{--}(1,0)_{--}]\\
\nonumber&\times U[(\lambda',\mu')(N_4,0)(\lambda,\mu)(0,N_4-1);(0,N_3-1)_{--}(1,0)_{--}]\Big\}\\
&\times\langle\sigma'\upsilon'\omega'S'\|\left[\left[a_{N_1}^{\dagger}\times a_{N_2}^{\dagger}\right]^{(\lambda_f,\mu_f)S_{f}}\times\left[\tilde{a}_{N_3-1}\times\tilde{a}_{N_4-1}\right]^{(\lambda,\mu) S_{i}}\right]^{\rho_{0}'(\lambda_0',\mu_0')S_0}\|\sigma\upsilon_{1}\omega_{1}S\rangle_{\rho'}.
\end{align}
The term $A_{1,0,0,1}$ contains TBDRMEs with $N_1$ increased by one and $N_4$ reduced by one:
\begin{align}
\nonumber&A_{1,0,0,1}=\frac{(-1)^{N_3+N_4-\lambda_{i}-\mu_i}}{6A}\sqrt{\frac{N_4(N_4+1)(N_4+2)(N_1+1)\dim(\lambda_i,\mu_i)}{\dim(0,N_3)}}\\
\nonumber&\times\sum_{(\lambda,\mu)(\lambda',\mu')(\lambda'',\mu'')\rho''}(-1)^{\lambda'+\mu'}\sqrt{\dim(\lambda',\mu')}U[(1,0)(N_1,0)(\lambda,\mu)(N_2,0);(N_1+1,0)_{--}(\lambda_f,\mu_f)_{--}]\\
\nonumber&\times U[(\lambda_i,\mu_i)(N_4,0)(\lambda',\mu')(0,N_4-1);(0,N_3)_{--}(1,0)_{--}]\\
\nonumber&\times U[(\lambda_0,\mu_0)(1,0)(\lambda_0',\mu_0')(1,0);(\lambda'',\mu'')_{--}(2,0)_{--}]\\
\nonumber&\times\bigg\{(-1)^{\lambda_{f}+\mu_f+\lambda_{0}'+\mu_0'-\lambda''-\mu''}\sqrt{\frac{\dim(\lambda_0,\mu_0)}{\dim(\lambda_f,\mu_f)}}U[(\lambda_0,\mu_0)(\mu_i,\lambda_i)(\lambda'',\mu'')(\lambda',\mu');(\lambda_f,\mu_f)\rho_0\rho''(1,0)_{--}]\\
\nonumber&\times U[(1,0)(\lambda_f,\mu_f)(\lambda_0',\mu_0')(\lambda',\mu');(\lambda,\mu)_{-}\rho_{0}'(\lambda'',\mu'')\rho''_{-}]\\
\nonumber&+(-1)^{\lambda+\mu}\sqrt{\frac{\dim(\lambda'',\mu'')}{\dim(\lambda,\mu)}}U[(1,0)(\lambda_f,\mu_f)(\lambda'',\mu'')(\lambda_i,\mu_i);(\lambda,\mu)_{-}\rho''(\lambda_0,\mu_0)\rho_{0-}]\\
\nonumber&\times U[(\lambda'',\mu'')(\mu_i,\lambda_i)(\lambda_0',\mu_0')(\lambda',\mu');(\lambda,\mu)\rho''\rho_{0}'(1,0)_{--}]\bigg\}\\
&\times\langle\sigma'\upsilon'\omega'S'\|\left[\left[a_{N_1+1}^{\dagger}\times a_{N_2}^{\dagger}\right]^{(\lambda,\mu) S_{f}}\times\left[\tilde{a}_{N_3}\times\tilde{a}_{N_4-1}\right]^{(\lambda',\mu')S_{i}}\right]^{\rho_{0}'(\lambda_0',\mu_0')S_0}\|\sigma\upsilon_{1}\omega_{1}S\rangle_{\rho'}.
\end{align}
The term $A_{0,1,0,1}$ contains TBDRMEs with $N_2$ increased by one and $N_4$ reduced by one:
\begin{align}
\nonumber&A_{0,1,0,1}=-\frac{(-1)^{\lambda_{f}+\mu_f+N_3+N_4-\lambda_{i}-\mu_i}}{6A}\sqrt{\frac{N_4(N_4+1)(N_4+2)(N_2+1)\dim(\lambda_i,\mu_i)}{\dim(0,N_3)}}\\
\nonumber&\times\sum_{(\lambda,\mu)(\lambda',\mu')(\lambda'',\mu'')\rho''}(-1)^{\lambda'+\mu'}\sqrt{\dim(\lambda',\mu')}U[(1,0)(N_2,0)(\lambda,\mu)(N_1,0);(N_2+1,0)_{--}(\lambda_f,\mu_f)_{--}]\\
\nonumber&\times U[(\lambda_i,\mu_i)(N_4,0)(\lambda',\mu')(0,N_4-1);(0,N_3)_{--}(1,0)_{--}]\\
\nonumber&\times U[(\lambda_0,\mu_0)(1,0)(\lambda_0',\mu_0')(1,0);(\lambda'',\mu'')_{--}(2,0)_{--}]\\
\nonumber&\times\bigg\{(-1)^{\lambda_{f}+\mu_f+\lambda_{0}'+\mu_0'-\lambda''-\mu''-\lambda-\mu}\sqrt{\frac{\dim(\lambda_0,\mu_0)}{\dim(\lambda_f,\mu_f)}}\\
\nonumber&\times U[(\lambda_0,\mu_0)(\mu_i,\lambda_i)(\lambda'',\mu'')(\lambda',\mu');(\lambda_f,\mu_f)\rho_0\rho''(1,0)_{--}]\\
\nonumber&\times U[(1,0)(\lambda_f,\mu_f)(\lambda_0',\mu_0')(\lambda',\mu');(\lambda,\mu)_{-}\rho_{0}'(\lambda'',\mu'')\rho''_{-}]\\
\nonumber&+\sqrt{\frac{\dim(\lambda'',\mu'')}{\dim(\lambda,\mu)}}U[(1,0)(\lambda_f,\mu_f)(\lambda'',\mu'')(\lambda_i,\mu_i);(\lambda,\mu)_{-}\rho''(\lambda_0,\mu_0)\rho_{0-}]\\
\nonumber&\times U[(\lambda'',\mu'')(\mu_i,\lambda_i)(\lambda_0',\mu_0')(\lambda',\mu');(\lambda,\mu)\rho''\rho_{0}'(1,0)_{--}]\bigg\}\\
&\times\langle\sigma'\upsilon'\omega'S'\|\left[\left[a_{N_1}^{\dagger}\times a_{N_2+1}^{\dagger}\right]^{(\lambda,\mu) S_{f}}\times\left[\tilde{a}_{N_3}\times\tilde{a}_{N_4-1}\right]^{(\lambda',\mu')S_{i}}\right]^{\rho_{0}'(\lambda_0',\mu_0')S_0}\|\sigma\upsilon_{1}\omega_{1}S\rangle_{\rho'}.
\end{align}
The term $A_{1,0,1,0}$ contains TBDRMEs with $N_1$ increased by one and $N_3$ reduced by one:
\begin{align}
\nonumber&A_{1,0,1,0}=-\frac{(-1)^{N_3+N_4}}{6A}\sqrt{\frac{N_3(N_3+1)(N_3+2)(N_1+1)\dim(\lambda_i,\mu_i)}{\dim(0,N_4)}}\\
\nonumber&\times\sum_{(\lambda,\mu)(\lambda',\mu')(\lambda'',\mu'')\rho''}\sqrt{\dim(\lambda',\mu')}U[(1,0)(N_1,0)(\lambda,\mu)(N_2,0);(N_1+1,0)_{--}(\lambda_f,\mu_f)_{--}]\\
\nonumber&\times U[(\lambda_i,\mu_i)(N_3,0)(\lambda',\mu')(0,N_3-1);(0,N_4)_{--}(1,0)_{--}]\\
\nonumber&\times U[(\lambda_0,\mu_0)(1,0)(\lambda_0',\mu_0')(1,0);(\lambda'',\mu'')_{--}(2,0)_{--}]\\
\nonumber&\times\bigg\{(-1)^{\lambda_{f}+\mu_f+\lambda_{0}'+\mu_0'-\lambda''-\mu''}\sqrt{\frac{\dim(\lambda_0,\mu_0)}{\dim(\lambda_f,\mu_f)}}U[(\lambda_0,\mu_0)(\mu_i,\lambda_i)(\lambda'',\mu'')(\lambda',\mu');(\lambda_f,\mu_f)\rho_0\rho''(1,0)_{--}]\\
\nonumber&\times U[(1,0)(\lambda_f,\mu_f)(\lambda_0',\mu_0')(\lambda',\mu');(\lambda,\mu)_{-}\rho_{0}'(\lambda'',\mu'')\rho''_{-}]\\
\nonumber&+(-1)^{\lambda+\mu}\sqrt{\frac{\dim(\lambda'',\mu'')}{\dim(\lambda,\mu)}}U[(1,0)(\lambda_f,\mu_f)(\lambda'',\mu'')(\lambda_i,\mu_i);(\lambda,\mu)_{-}\rho''(\lambda_0,\mu_0)\rho_{0-}]\\
\nonumber&\times U[(\lambda'',\mu'')(\mu_i,\lambda_i)(\lambda_0',\mu_0')(\lambda',\mu');(\lambda,\mu)\rho''\rho_{0}'(1,0)_{--}]\bigg\}\\
&\times\langle\sigma'\upsilon'\omega'S'\|\left[\left[a_{N_1+1}^{\dagger}\times a_{N_2}^{\dagger}\right]^{(\lambda,\mu)S_{f}}\times\left[\tilde{a}_{N_3-1}\times\tilde{a}_{N_4}\right]^{(\lambda',\mu')S_{i}}\right]^{\rho_{0}'(\lambda_0',\mu_0')S_0}\|\sigma\upsilon_{1}\omega_{1}S\rangle_{\rho'}.
\end{align}
The term $A_{0,1,1,0}$ contains TBDRMEs with $N_2$ increased by one and $N_3$ reduced by one:
\begin{align}
\nonumber A_{0,1,1,0}=&\frac{(-1)^{\lambda_{f}+\mu_f+N_3+N_4}}{6A}\sqrt{\frac{N_3(N_3+1)(N_3+2)(N_2+1)\dim(\lambda_i,\mu_i)}{\dim(0,N_4)}}\\
\nonumber&\times\sum_{(\lambda,\mu)(\lambda',\mu')(\lambda'',\mu'')\rho''}\sqrt{\dim(\lambda',\mu')}U[(1,0)(N_2,0)(\lambda,\mu)(N_1,0);(N_2+1,0)_{--}(\lambda_f,\mu_f)_{--}]\\
\nonumber&\times U[(\lambda_i,\mu_i)(N_3,0)(\lambda',\mu')(0,N_3-1);(0,N_4)_{--}(1,0)_{--}]\\
\nonumber&\times U[(\lambda_0,\mu_0)(1,0)(\lambda_0',\mu_0')(1,0);(\lambda'',\mu'')_{--}(2,0)_{--}]\\
\nonumber&\times\bigg\{(-1)^{\lambda_{f}+\mu_f+\lambda_{0}'+\mu_0'-\lambda''-\mu''-\lambda-\mu}\sqrt{\frac{\dim(\lambda_0,\mu_0)}{\dim(\lambda_f,\mu_f)}}\\
\nonumber&\times U[(\lambda_0,\mu_0)(\mu_i,\lambda_i)(\lambda'',\mu'')(\lambda',\mu');(\lambda_f,\mu_f)\rho_0\rho''(1,0)_{--}]\\
\nonumber&\times U[(1,0)(\lambda_f,\mu_f)(\lambda_0',\mu_0')(\lambda',\mu');(\lambda,\mu)_{-}\rho_{0}'(\lambda'',\mu'')\rho''_{-}]\\
\nonumber&+\sqrt{\frac{\dim(\lambda'',\mu'')}{\dim(\lambda,\mu)}}U[(1,0)(\lambda_f,\mu_f)(\lambda'',\mu'')(\lambda_i,\mu_i);(\lambda,\mu)_{-}\rho''(\lambda_0,\mu_0)\rho_{0-}]\\
\nonumber&\times U[(\lambda'',\mu'')(\mu_i,\lambda_i)(\lambda_0',\mu_0')(\lambda',\mu');(\lambda,\mu)\rho''\rho_{0}'(1,0)_{--}]\bigg\}\\
&\times\langle\sigma'\upsilon'\omega'S'\|\left[\left[a_{N_1}^{\dagger}\times a_{N_2+1}^{\dagger}\right]^{(\lambda,\mu)S_{f}}\times\left[\tilde{a}_{N_3-1}\times\tilde{a}_{N_4}\right]^{(\lambda',\mu')S_{i}}\right]^{\rho_{0}'(\lambda_0',\mu_0')S_0}\|\sigma\upsilon_{1}\omega_{1}S\rangle_{\rho'}.
\end{align}
Finally, the term $A_{1,1,0,0}$ contains TBDRMEs with $N_1$ and $N_2$ increased by one:
\begin{align}
\nonumber A_{1,1,0,0}=&\frac{1}{A}\sqrt{\frac{(N_1+1)(N_2+1)}{2}}\\
\nonumber&\times\sum_{(\lambda,\mu)(\lambda',\mu')}(-1)^{\lambda'+\mu'+\lambda_{0}'+\mu_0'}U[(2,0)(\lambda_f,\mu_f)(\lambda_0',\mu_0')(\lambda_i,\mu_i);(\lambda,\mu)_{-}\rho_{0}'(\lambda_0,\mu_0)\rho_{0-}]\\
\nonumber&\times U[(\lambda_f,\mu_f)(1,0)(\lambda,\mu)(1,0);(\lambda',\mu')_{--}(2,0)_{--}]\\
\nonumber&\times\Big\{(-1)^{\lambda_{f}+\mu_f}U[(1,0)(N_2,0)(\lambda',\mu')(N_1,0);(N_2+1,0)_{--}(\lambda_f,\mu_f)_{--}]\\
\nonumber&\times U[(1,0)(N_1,0)(\lambda,\mu)(N_2+1,0);(N_1+1,0)_{--}(\lambda',\mu')_{--}]\\
\nonumber&+(-1)^{\lambda+\mu}U[(1,0)(N_1,0)(\lambda',\mu')(N_2,0);(N_1+1,0)_{--}(\lambda_f,\mu_f)_{--}]\\
\nonumber&\times U[(1,0)(N_2,0)(\lambda,\mu)(N_1+1,0);(N_2+1,0)_{--}(\lambda',\mu')_{--}]\Big\}\\
&\times\langle\sigma'\upsilon'\omega'S'\|\left[\left[a_{N_1+1}^{\dagger}\times a_{N_2+1}^{\dagger}\right]^{(\lambda,\mu)S_{f}}\times\left[\tilde{a}_{N_3}\times\tilde{a}_{N_4}\right]^{(\lambda_i,\mu_i)S_{i}}\right]^{\rho_{0}'(\lambda_0',\mu_0')S_0}\|\sigma\upsilon_{1}\omega_{1}S\rangle_{\rho'}\Bigg]\Bigg\}.
\end{align}
The relation~(\ref{recurrence2}) relates the TBDRMEs between the symplectic many-body basis states with TBDRMEs between basis states with numbers of HO quanta reduced by two. Therefore it can, along with the conjugation relation
\begin{align}\label{con2}
\nonumber\langle\sigma'\upsilon'\omega'S'&\|\left[\left[a^{\dagger}_{N_1}\times a^{\dagger}_{N_2}\right]^{(\lambda_f,\mu_f)S_{f}}\times\left[\tilde{a}_{N_3}\times\tilde{a}_{N_4}\right]^{(\lambda_i,\mu_i)S_{i}}\right]^{\rho_0(\lambda_0,\mu_0)S_0}\|\sigma\upsilon\omega S\rangle_{\rho}\\
\nonumber=&(-1)^{\lambda_{f}+\mu_f+\lambda_{i}+\mu_i-\lambda_0-\mu_0+N_1+N_2+N_3+N_4+\lambda_{\omega'}+\mu_{\omega'}-\lambda_{\omega}-\mu_{\omega}+S-S'+\rho_{0,\textnormal{max}}}\\
\nonumber&\times\sqrt{\frac{(2S+1)\dim\omega}{(2S'+1)\dim\omega'}}\sum_{\rho_0'}(-1)^{\rho_0'}\Phi_{\rho_0\rho_0'}[(\lambda_f,\mu_f),(\lambda_i,\mu_i);(\lambda_0,\mu_0)]\\
&\times \langle\sigma\upsilon\omega S\|\left[\left[a^{\dagger}_{N_4}\times a^{\dagger}_{N_3}\right]^{(\mu_i,\lambda_i)S_{i}}\times\left[\tilde{a}_{N_2}\times\tilde{a}_{N_1}\right]^{(\mu_f,\lambda_f)S_{f}}\right]^{\rho_0'(\mu_0,\lambda_0)S_0}\|\sigma'\upsilon'\omega'S'\rangle_{\rho}^{*},
\end{align}
where $\rho_{0,\rm max}$ is the outer multiplicity of the $\grpsu{3}$ coupling $(\lambda_f,\mu_f)\times(\lambda_i,\mu_i)\to(\lambda_0,\mu_0)$, be used to calculate the TBDRMEs recursively from the TBDRMEs between the LGIs.

The correctness of the recurrence relations~(\ref{recurrence1}) and~(\ref{recurrence2}) was numerically verified for all OBDRMEs (TBDRMEs) in the model space of states of $^6$Li carrying up to six (four) excitation HO quanta. The verification was done by also calculating these same density matrix elements by the alternative nonrecursive, but highly inefficient, method of explicitly expanding the symplectic many-body basis states in terms of $\grpu{3}$ many-body basis states, and then calculating the OBDRMEs and TBDRMEs from the OBDRMEs and TBDRMEs between these $\grpu{3}$ many-body basis states. The calculations were done with the use of the computer codes \texttt{LSU3shell}~\cite{dytrych2016} and \texttt{spncci}~\cite{spncci}.

\section{Conclusion}

We derived recurrence relations for calculation of one-body and two-body density matrix elements between symplectic many-body basis states.  (The principles of the derivation could be applied to higher-body density matrix elements as well, allowing for calculation of matrix elements of an arbitrary many-body operator, although the resulting recurrence relations would be more complex.)

Specifically, we work with the intrinsic realization of $\grpsptr$, such that the many-body space is free of spurious center-of-mass excitation, and we fully account for the center-of-mass correction to the symplectic raising operator in deriving the recurrence.  The correctness of our recurrence relations was tested by numerically benchmarking to results obtained using a largely independent, but computationally much less efficient, approach, based on direct expansion in a $\grpu{3}$ basis.

Such one-body and two-body density matrix elements between the symplectic many-body basis states can be used to calculate matrix elements of arbitrary one-body and two-body operators in the SpNCCI many-body basis, provided the one-body or two-body matrix elements of these operators are available. The density matrix elements can also be useful for integrating the SpNCCI framework with other many-body methods, such as the normal ordered two-body approximation of a three-body interaction, the in-medium similarity renormalization group method, and the resonating-group method.

\begin{acknowledgments}
We thank Anna E.~McCoy for valuable discussions and suggestions.
This material is based upon work supported by the U.S.~Department of Energy,
Office of Science, Office of Nuclear Physics, under Award Number
DE-FG02-95ER-40934.  TRIUMF receives federal
funding via a contribution agreement with the National Research Council of
Canada.
\end{acknowledgments}

\bibliographystyle{apsrev4-2}
\nocite{control:title-on}

\end{document}